\documentclass[11pt,a4paper,twoside,groupcitations]{article}
\usepackage[T1]{fontenc}
\usepackage[ansinew]{inputenc}
\usepackage[english]{babel}
\usepackage{amsfonts}
\usepackage{amsmath}
\usepackage{bm}
\usepackage{array,upgreek,accents}
\usepackage{amsthm}
\usepackage{amssymb}
\usepackage{graphicx}
\usepackage{braket}
\usepackage{eucal}
\usepackage{verbatim}
\usepackage[table]{xcolor}
\usepackage{caption}
\usepackage{cite}
\usepackage{textcomp}
\usepackage{float}
\usepackage{multicol}
\usepackage{tikz}
\usetikzlibrary{positioning,arrows}
\usetikzlibrary{decorations.pathmorphing}
\usetikzlibrary{decorations.markings}
\usetikzlibrary{calc,decorations.markings}
\usetikzlibrary{arrows,shapes}
\usetikzlibrary{matrix,arrows}

\usepackage[colorlinks,hyperindex,unicode]{hyperref}
\definecolor{green1}{RGB}{0,128,0} 
\hypersetup{hidelinks,colorlinks=true,breaklinks=true,urlcolor= blue}
\hypersetup{%
colorlinks = true,
linkcolor  = blue,
citecolor = cyan,
}
\newcommand{\beq}{\begin{eqnarray}}
\newcommand{\eeq}{\end{eqnarray}}
\newcommand{\be}{\begin{eqnarray}}
\newcommand{\ee}{\end{eqnarray}}
\newcommand{\pro}[2]{\mbox{$\langle\, #1 \mid #2\,\rangle$}}
\newcommand{\expec}[1]{\mbox{$\langle\, #1\,\rangle$}}

\renewcommand{\d}{\mbox{${\rm d}$}} 
\newcommand{\lp}{\ell_{\rm p}}
\newcommand{\mpl}{m_{\rm p}}
\newcommand{\gn}{G_{\rm N}}

\newcommand{\Rh}{R_{\rm H}}

\newcommand{\dd }{{\mathrm d}}
\newcommand{\tb}{\color{black}}

\usepackage{caption}
\title{\bf  Black holes with a charged quantum dust core}
\author{R.~Casadio$^{ab}$\thanks{E-mail: casadio@bo.infn.it},
$\ $
R.~da~Rocha$^{c}$\thanks{E-mail: roldao.rocha@ufabc.edu.br},
$\ $
A.~Giusti$^{d}$\thanks{E-mail: A.Giusti@sussex.ac.uk}
$\ $
and
P.~Meert$^{e}$\thanks{E-mail: pedro.meert@unesp.br}
\\
\\
$^a${\em Dipartimento di Fisica e Astronomia, Universit\`a di Bologna}
\\
{\em via Irnerio~46, 40126 Bologna, Italy}
\\
\\
$^b${\em I.N.F.N., Sezione di Bologna, I.S.~FLAG}
\\
{\em viale B.~Pichat~6/2, 40127 Bologna, Italy}
\\
\\
$^c${\em Federal University of ABC, Center of Mathematics}
\\
{\em Santo Andr\'e, 09210-580, Brazil}
\\
\\
$^d$ {\em Department of Physics and Astronomy}
\\
{\em University of Sussex, Brighton, BN1~9QH, United Kingdom}
\\
\\
$^e${\em Instituto de F\'isica Te\'orica, Unesp,}
\\
{\em S\~ao Paulo, 01140-070, Brazil}
}
\begin{document}
\maketitle
\begin{abstract}
To understand the nature of the black holes that exist in the Universe, it is also
necessary to study what happens to the (quantum) matter that collapses and forms such objects.
In this work, we consider a dust ball with an electrically charged central core and study its quantum
spectrum by quantising the geodesic equation for individual dust particles in the corresponding
Reissner-Nordstr\"om spacetime.
As in the neutral case investigated previously, we find a ground state of the dust ball
with the size of a fraction of the outer horizon.
Moreover, we determine a self-consistent configuration of layers in the ground state
corresponding to an effective mass function that increases linearly with the areal radius
and has no inner Cauchy horizon.
We then briefly speculate on the possible phenomenological consequences
for the endpoint of the gravitational collapse.
\end{abstract}
\newpage
\section{Introduction}
\label{sec:intro}
\setcounter{equation}{0}
All of the known black hole solutions of General Relativity are characterised by the
Arnowitt-Deser-Misner (ADM)~\cite{ADM} mass $M$, electric charge $Q$, and angular momentum
$J$, and hide classical spacetime singularities inside the event horizon~\cite{Wald:1984rg}.
The existence of the horizon further implies that the spacetime is geodesically incomplete~\cite{HE},
which hints to possible violations of causality in the matter dynamics.
The overall picture is particularly simple in the spherically symmetric Schwarzschild geometry,
which essentially results from neglecting the matter that has collapsed to form it~\cite{geroch,balasin},
and the corresponding causal structure only contains the event horizon that hides the central
singularity where certain geometric invariants built from the Riemann tensor diverge.
\par
Non-trivial internal structures emerge as soon as one adds some matter source, like the electric field
in the spherically symmetric Reissner-Nordstr\"om spacetime.
In fact, beside the central singularity and outer event horizon, there now appears an inner Cauchy horizon,
which was shown to cause potentially strong instabilities already long
ago~\cite{McNamara,Gursel:1979zza,chandra,Poisson:1989zz}
(see also Refs.~\cite{Gnedin:1993nau,Brady:1995ni,Burko:1997zy,Marolf:2011dj,
Eilon:2016osg,Carballo-Rubio:2024dca,Balbinot:2023grl}
for some more recent results).
Since all of these issues are explicitly connected with the (inevitable) presence and behaviour of matter,
one would hope that they can be resolved by properly taking into account the quantum nature
of matter (and gravity), which has in fact been proposed in several approaches (see,
{\em e.g.}~Refs.~\cite{Casadio:2019tfz,Casadio:1998yr,Kuntz:2019lzq,Kuntz:2019gup,
Haggard:2014rza,Bonanno:2023rzk}).
\par
One of the few cases in which the gravitational collapse leading to black hole formation can be studied
analytically is given by the Lema\^itre-Tolman-Bondi model~\cite{Lemaitre:1927zz,Tolman:1934za,Bondi:1947fta}
of a dust ball with ADM mass $M$ and areal radius $r=R_{\rm s}(\tau)$, where $\tau$ is the
proper time at the surface.
Dust particles are not subjected to any pressure and follow geodesics in their own
geometry~\cite{EIH}, so that the density profile~\footnote{In particular, the density is homogeneous
in the Oppenheimer-Snyder~\cite{Oppenheimer:1939ue} model.}
shrinks during the collapse without changing shape.
Many studies of this model start from a reduction of degrees of freedom
based on spherical symmetry and continuity of the fluid used to describe the dust,
which uniquely determine the interior Lema\^itre-Tolman-Bondi metric and the exterior Schwarzschild metric.
From the Einstein-Hilbert action for such metrics (and the proper junction conditions at the surface),
one can identify a few collective degrees of freedom, including the ball radius $R_{\rm s}$,
which can be canonically quantised (see, {\em e.g.}~Refs.~\cite{Kiefer:2004xyv,Casadio:1998ta,
Vaz:2011zz,Kiefer:2019csi,Piechocki:2020bfo,Schmitz:2020vdr,Husain:2022gwp,Giesel:2022rxi}).
\par
The above approach to quantisation dispenses with the fact that astrophysical objects with enough mass
to form a black hole must contain a huge number of matter particles.~\footnote{$M\simeq 10^{57}\,\mu$
for a solar mass object made of neutrons.}
Moreover, the classical Einstein field equations are analogous to thermodynamical laws~\cite{Jacobson:1995ab}
which, for black holes~\cite{Bardeen:1973gs}, appear to suggest an even larger number of gravitational
excitations~\cite{bekenstein}.
It seems therefore more appropriate to describe {\em a priori\/} a ball of dust  with the quantum state
for a very large number of particles of mass $\mu$ and derive a collective description {\em a posteriori\/}.
This is the alternative viewpoint advocated in Refs.~\cite{Casadio:2021cbv,Casadio:2023ymt} for
studying the case of dust without electric fields.
\par
In order to preserve the (approximate) spherical symmetry, all particles are assumed to move
radially without crossing paths and their number is large enough to form an ``almost'' continuous distribution.
Moreover, since $\mu\ll M$, the back-reaction of individual particles on the local geometry is neglected
by considering a course-grained layering of the ball, in which $r=R_i(\tau)$ denotes trajectories
on the inner border of the $i^{\rm th}$ layer~\cite{Casadio:2023ymt}.
Between the surfaces $r=R_i(\tau)$ and $r=R_{i+1}(\tau)$ should lie enough intermediate
trajectories that the mass contribution of the particles at $r=R_i(\tau)$ is negligible with respect
to the total mass of the $\nu_i\gg 1$ particles in the whole $i^{\rm th}$ layer defined by
$R_i\le r<R_{i+1}$.
Birkhoff's theorem then implies that particles following a trajectory with $r=R_i$ are not
affected by the particles at $r= R_{j>i}$, but only by those at $r=R_{j<i}$, whose number
and ADM mass do not change in time.
Trajectories of dust particles (at the inner border of a given layer) are individually quantised
and a condition is imposed to ensure that the fuzzy quantum layers defined by the positions
of these particles remain orderly nested in the global quantum ground state in agreement
with the uncertainty principle (see Ref.~\cite{Casadio:2023ymt} and Section~\ref{sec:QRN_core}
for all the details).
\par
Once the wavefunction $\psi=\psi(r)$ for the ground state is obtained, an (effective) Misner-Sharp-Hernandez
mass function~\cite{Misner:1964je,Hernandez:1966zia} for the core can be defined by
\be
m(r)
\equiv
4\,\pi
\int_0^r
\rho(x)\,x^2\,\d x
\sim
4\,\pi
\int_0^r
|\psi(x)|^2\,x^2\,\d x
<
\infty
\ ,
\qquad
{\rm for}
\ r>0
\ ,
\label{Qcond}
\ee
where $\rho=\rho(r)$ is the effective energy density~\cite{Casadio:2023ymt}.
In particular, one finds that $m(r\to\infty)=M$ and
\be
\rho\sim r^{-2}
\quad
{\rm and}
\quad
m\sim r
\ ,
\qquad
{\rm for}
\ r\to 0
\ ,
\label{m~r}
\ee
which ensures that $m(0)=0$ and the central curvature singularity of the vacuum Schwarzschild
geometry~\cite{HE} is replaced by an integrable
singularity~\cite{Casadio:2021eio,Casadio:2022ndh,Casadio:2023iqt}.
The centre of the core is, therefore, a region where the curvature invariants and the effective
energy-momentum tensor diverge, but their volume integrals and tidal forces acting on radial
geodesics remain finite~\cite{Lukash:2013ts}.
\par
It is well known that singularities which plague black hole solutions of the vacuum Einstein
equations~\cite{HE} can also be removed by imposing regularity conditions on the (effective)
energy density and scalar invariants inspired by classical physics~\cite{Carballo-Rubio:2023mvr}.
However, this procedure usually induces the existence of an inner Cauchy horizon, which is instead
not the case for a mass function of the form in Eq.~\eqref{m~r}~\cite{Casadio:2023iqt}.
As we recalled above, adding an electric field is the simplest way to induce
the presence of a Cauchy horizon in the (otherwise empty) Reissner-Nordstr\"om spacetime.
It is therefore interesting to further study the possible existence of inner horizons inside
black holes by extending the previous investigations of the dust
ball~\cite{Casadio:2021cbv,Casadio:2023ymt} to include an electric charge $Q$.
\par
{\tb
The static and spherically symmetric line element
\begin{equation}
\label{eq:rn}
\dd s^{2}
=
-f(r)\,\dd t^{2}
+
\frac{\dd r^{2}}{f(r)}
+
r^{2}\,\dd\Omega^{2}
\end{equation}
represents the vacuum Reissner-Nordstr\"om spacetime generated by a source of ADM mass $M$
and charge $Q$ for
\be
f\left( r \right)
=
f_{\rm RN}
=
1
-
\frac{2\,\gn\,M}{r}
+
\frac{\gn\,Q^{2} }{ r^{2} }
\ ,
\label{fQ}
\ee
where $G_N$ is Newton's constant.
The metric function in Eq.~\eqref{fQ} can have two (possibly degenerate) zeroes, associated with
horizons, located at}
\be
\label{eq:RN-hor}
R_\pm
=
\gn\,M
\pm
\sqrt{\gn^2\,M^2-\gn\,Q^2}
\ ,
\ee
provided
\be
Q^2\le \gn\,M^2
\ .
\label{condR+-}
\ee
In particular, the sphere $r=R_+$ is the outer event horizon and $r=R_-$ is a Cauchy horizon~\cite{HE}.
\par
In the present work, we shall consider the simplest case of a dust ball of ADM mass $M$ with
the charge $Q$ localised inside a spherical innermost core of mass $\mu_0=\epsilon_0\,M$
and radius $r=R_1$.
This core is surrounded by a number $N\ge 1$ of electrically neutral layers of inner radius $r=R_i$,
thickness $\Delta R_i=R_{i+1}-R_i$, and mass $\mu_i=\epsilon_i\,M$, where $\epsilon_i$ is the fraction
of ADM mass associated with the $\nu_i$ dust particles in the $i^{\rm th}$ layer.
The gravitational mass inside the ball $r<R_i$ will be denoted by
\be
M_i
=
\sum_{k=0}^{i-1}
\mu_k
=
M\,\sum_{k=0}^{i-1}\epsilon_k
\ ,
\ee
with $M_1=\mu_0$ and $M_{N+1}=M$.
The radius $R_1$ and the mass $M_1=\mu_0$ of the innermost core, as well as the thickness $\Delta R_i$
of each layer, can take arbitrarily small values in the classical picture, for example by increasing the number
$N$ of layers.
{\tb In this configuration, dust particles on the inner surface of the $i^{\rm th}$ layer will move along
radial geodesics (paramaterized by the proper time $\tau$), $r=R_i(\tau)$, of the Reissner-Nordstr\"om
spacetime~\eqref{eq:rn} with}
\begin{equation}
\label{eq:rn-metric}
f \left( r \right)
=
f_i
=
1 - \frac{2\,\gn\,M_i}{r}
+
\frac{\gn\,Q^{2} }{ r^{2}}
\end{equation}
In particular, the mass-shell condition for the 4-velocity of components
$u_i^\mu=\dd x_i^\mu/\dd\tau=(\dot t_i,\dot R_i,0,0)$ yields the Hamiltonian constraint equation
\begin{equation}
\label{eq:geodesic}
H_i
=
\frac{P_i^{2}}{2\,\mu}-\frac{\gn\,\mu\,M_i}{R_i}+\frac{\gn\,\mu\,Q^{2}}{2\,R_i^{2}}
=
\frac{\mu}{2}\left(\frac{E_i^{2}}{\mu^{2}}-1\right)
\equiv
\varepsilon_i
\ ,
\end{equation}
where $P_i=\mu\,\dot R_i$ is the momentum conjugated to $r=R_i(\tau)$, and $E_i$ is the conserved momentum
conjugated to $t_i=t_i(\tau)$.~\footnote{The conserved angular momentum conjugated to $\phi_i=\phi_i(\tau)$
vanishes for purely radial motion.}
It is important to remark that we are assuming the mass $\mu\ll M_i$ for all $i=0,\ldots,N+1$ or,
equivalently, the numbers $\nu_i\gg 1$, so that individual dust particles can be described as test
particles with a good approximation.
\par
Eq.~\eqref{eq:geodesic} can be canonically quantised, similarly to the equation for the
electron's trajectory in the quantum mechanical treatment of the hydrogen atom, and a spectrum
of bound states will be found like in the neutral case~\cite{Casadio:2021cbv,Casadio:2023ymt}.
We will first analyse the bound states for dust particles at the surface of the ball following
Ref.~\cite{Casadio:2021cbv} in the next Section;
a more refined description of the interior will then be obtained in Section~\ref{sec:mlayer_core}
by considering multiple layers like in Ref.~\cite{Casadio:2023ymt};
concluding remarks and outlook will be given in Section~\ref{sec:conc}.
\section{Dust ball with charged inner core}
\label{sec:QRN_core}
\setcounter{equation}{0}
Dust particles on the surface of the ball of radius $r=R_{N+1}(\tau)\equiv R_{\rm s}$ will fall radially
in the Reissner-Nordstr\"om metric defined by Eq.~\eqref{eq:rn-metric} with $M_{N+1}=M$
and their geodesic motion will be described by the Hamiltonian constraint
\begin{equation}
\label{eq:geodesic_gen}
H
=
\frac{P_{\rm s}^{2}}{2\,\mu}-\frac{\gn\,\mu\,M}{R_{\rm s}}+\frac{\gn\,\mu\,Q^{2}}{2\,R_{\rm s}^{2}}
=
\frac{\mu}{2}\left(\frac{E^{2}}{\mu^{2}}-1\right)
=
\varepsilon
\ ,
\end{equation}
where $P_{\rm s}=\mu\,\dot R_{\rm s}$ is the momentum conjugated to $R_{\rm s}=R_{\rm s}(\tau)$
and $E$ is the conserved momentum conjugated to $t=t_{\rm s}(\tau)$.
\par
Canonical quantisation is obtained by replacing $P_{\rm s}\mapsto \hat P_{\rm s}=-i\,\hbar\,\partial/\partial{R_{\rm s}}$
and, after some manipulations, the time-independent Schr\"odinger-like equation
\be
\hat{H}\,\Psi
=
\varepsilon\,\Psi
\ee
can be written as the generalised associated Laguerre equation~\footnote{We will often use units
with $\gn = \lp/\mpl $ and $\hbar = \lp\,\mpl$, where $\lp$ is the Planck length and $\mpl$ the Planck mass.}
\begin{equation}
\label{eq:wavefunction-ode}
\left(
\frac{\dd^{2}}{\dd x^{2}}
+
\frac{2\,\mu^{2}\, M}{\gamma\, \lp\, \mpl^{3}\,x}
-
\frac{\mu^{2}\,Q^{2}}{\lp\,\mpl^{3}\,x^2}
-\frac{1}{4}
\right)
\Psi
=
0
\ ,
\end{equation}
where $x=\gamma\,R_{\rm s}$ with
\begin{equation}
\label{eq:nk-gamma}
\gamma^{2}
=
-\frac{8\,\mu\,\varepsilon}{\mpl^{2}\,\lp^{2}}
\ .
\end{equation}
Orthonormal solutions, in the scalar product
\be
\pro{\psi}{\chi}
=
4\,\pi\int_0^\infty
\psi^*(R_{\rm s})\,\chi(R_{\rm s})\,R_{\rm s}^2\,\dd R_{\rm s}
\ ,
\label{product}
\ee
{\tb are then given by the wavefunctions
\begin{equation}
\label{eq:wavefunction-sol}
\Psi_{n\alpha}(R_{\rm s})
=
A_{n\alpha}\,
e^{-\frac{\gamma}{2}\,R_{\rm s}}\,
r^{\frac{\alpha-1}{2}}\,L_{n-1}^{\alpha}(\gamma\,R_{\rm s})
\ ,
\end{equation}
where $L^\alpha_{n-1}$ are generalised Laguerre polynomials of integer order
$n\geq 1$ and continuous parameter
\begin{equation}
\label{eq:k2-Q}
\alpha^{2}
=
1+
\frac{4\,\mu^{2}\,Q^{2}}{\lp\,\mpl^{3}}
\ ,
\end{equation}
from which one can see that $|\alpha|\ge 1$.
The normalisation reads
\begin{equation}
\label{eq:wavefunction-norm}
A_{n\alpha}^{2}
=
\frac{\gamma^{\alpha+2}\,\Gamma(n)}
{4\,\pi^{2}\,(2\,n+\alpha-1)\,\Gamma^3(\alpha+n)}
\ ,
\end{equation}
where $\Gamma$ is the Euler gamma function.
By introducing
\begin{equation}
\label{eq:nk-int}
\beta_{n\alpha}
\equiv
n+\frac{\alpha-1}{2}
=
\frac{2\,\mu^{2}\, M}{\gamma\, \lp\, \mpl^{3}}
\ ,
\end{equation}
the corresponding quantised energy spectrum $\varepsilon=\varepsilon_{n\alpha}$
can be obtained by solving Eqs.~\eqref{eq:nk-int} and~\eqref{eq:nk-gamma} for $\gamma$,
and reads
\begin{equation}
\label{qes}
\varepsilon_{n\alpha}
=
-\frac{2\,\mu^{3}\,M^{2}}{\left( 2\,n+\alpha-1 \right)^{2}\mpl^{4}}
=
-\frac{ \mu^{3}\,M^{2} }{2\,\beta_{n\alpha}^{2}\, \mpl^{4}}
\ ,
\end{equation}
which is discrete and bounded below.}
This implies the physically expected fact that an infinite amount of energy cannot
be extracted from the system, in line with other quantum-mechanical descriptions
of the Reissner-Nordstr\"om black holes, such as the Hamiltonian quantum theory
of spherically symmetric and asymptotically flat electrovacuum spacetimes in
Ref.~\cite{Makela:1997rx}.
\par
The wavefunctions~\eqref{eq:wavefunction-sol} can now be written as
\be
\label{eq:wavefunction-n}
\pro{R_{\rm s}}{n\alpha}
=
\Psi_{n\alpha}(R_{\rm s})
=
A_{n\alpha}\,
\exp\!\left(-\frac{\mu^{2}\,M\,R_{\rm s}}{\beta_{n\alpha}\,\lp\,\mpl^{3}}\right)
R_{\rm s}^{\frac{\alpha-1}{2}}\,
L_{n-1}^{\alpha}\!\left(\frac{2\,\mu^{2}\,M\,R_{\rm s}}{\beta_{n\alpha}\,\lp\,\mpl^{3}}\right)
\ ,
\ee
where the integer $n\ge 1$ and the value of $ \alpha $ is determined from the charge
of the system according to Eq.~\eqref{eq:k2-Q}.
We also note that the quantised energy spectrum~\eqref{qes} is similar to the findings
of Ref.~\cite{Berezin:1996ju}, as it supports the existence of a lowest energy level.
However, unlike the spectrum of the hydrogen atom, the quantised energy spectrum~\eqref{qes}
depends not only on the charge $Q$ and mass $\mu$ of the dust particles but also on
the total ADM mass $M$ of the ball.
\subsection{Ball radius and uncertainty}
\label{ssec:expect-values}
{\tb The wavefunction~\eqref{eq:wavefunction-n} determines the probability distribution [in the
scalar product~\eqref{product}] for dust particles on the shell at the surface of the ball to be found
at the radial position $R_{\rm s}$ given their mass $\mu$ and the ball ADM mass $M$,
with $n$ and $\alpha$ determined by $M$ and the charge $Q$ of the system.
The overall size of the dust ball is therefore just given by the expectation value
\begin{equation}
\label{eq:expct-r}
\bra{n\alpha}\hat R_{\rm s}\ket{n\alpha}
=
\frac{12\,\beta_{n\alpha}^{2}-\alpha^{2}+1}{4\,\beta_{n\alpha}\,\gamma_{n\alpha}}
\ ,
\end{equation}
where we used $\hat R_{\rm s}\ket{n\alpha}=R_{\rm s}\ket{n\alpha}$ and defined}
\be
\label{eq:gamma-na}
\gamma_{n\alpha}^2
=
-\frac{8\,\mu\,\varepsilon_{n\alpha}}{\mpl^{2}\,\lp^{2}}
=
\frac{4\,\mu^4\,M^2}{\beta^2_{n\alpha}\,\lp^2\,\mpl^6}
\ .
\ee
Taking $\gamma_{n\alpha}>0$ yields
\be
\bra{n\alpha}\hat R_{\rm s}\ket{n\alpha}
=
\frac{\lp\,\mpl^3}{8\,\mu^2\,M}\left(12\,\beta_{n\alpha}^{2}-\alpha^{2}+1\right)
\ .
\ee
\par
{\tb It is also important to estimate the quantum uncertainty for the size of the ball.
From
\begin{equation}
\label{eq:expct-r2}
\bra{n\alpha}\hat R_{\rm s}^2\ket{n\alpha}
=
\frac{20\,\beta_{n\alpha}^{2}-3\,\alpha^{2}+7}{2\,\gamma_{n\alpha}^{2}}
\ ,
\end{equation}
and the definition of the variance $\Delta R_{\rm s}\equiv\sqrt{\expec{\hat R_{\rm s}^{2}}-\expec{\hat R_{\rm s}}^2}$,
we obtain the ratio
\be
\label{eq:r-var}
\frac{\Delta R_{\rm s}}{\bra{n\alpha}\hat R_{\rm s}\ket{n\alpha}}
=
\frac{\sqrt{\bra{n\alpha}\hat R_{\rm s}^2\ket{n\alpha}-\bra{n\alpha}\hat R_{\rm s}\ket{n\alpha}^2}}
{\bra{n\alpha}\hat R_{\rm s}\ket{n\alpha}}
=
\frac{\sqrt{16\,\beta_{n\alpha}^2\left(\beta_{n\alpha}^{2}+2\right)-\left(\alpha^{2}-1\right)^{2}}}
{12\,\beta_{n\alpha}^{2}-\alpha^{2}+1}
\ .
\ee
which is a measure of how fuzzy the dust ball is in the state $\ket{n\alpha}$.
The relevance of this result will become clearer in the following.}
\subsection{Ground state}
\label{ssec:gs-wavefunction}
In General Relativity, the conserved momentum $E^2\ge 0$, which carries on to the quantum
theory by constraining (from below) the possible values of the quantum number
$n$~\cite{Casadio:2021cbv,Casadio:2023ymt}.
{\tb  In particular, the ground state is defined by the minimum value $E^2=0$, which is equivalent
to $\varepsilon=-\mu/2$ from the right-most equality in Eq.~\eqref{eq:geodesic_gen}.
The corresponding value of $\gamma$ can be determined from Eq.~\eqref{eq:nk-gamma},
namely
\be
\gamma_{n\alpha}
=
\frac{2\,\mu}{\mpl\,\lp}
\equiv
\gamma_\mu
\ ,
\label{g_mu}
\ee
and leads to the minimum value of
\be
\label{eq:n-gs}
\beta_{n\alpha}
=
\frac{\mu\,M}{\mpl^{2}}
\equiv
\beta_M
\ .
\ee
The integer quantum number $n$ corresponding to the ground state can finally be expressed
in terms of the ADM mass $M$, dust particle mass $\mu$, and charge $Q$ by substituting
Eqs.~\eqref{eq:n-gs} and~\eqref{eq:k2-Q} into Eq.~\eqref{eq:nk-int}, to wit}
\begin{equation}
\label{eq:n-integer}
n
=
\frac{\mu\, M}{\mpl^{2}}
+
\frac{1}{2}
\left(
1-
\sqrt{1+ \frac{4\,Q^{2}\,\mu^{2}}{\mpl^{3}\,\lp}}
\right)
\equiv
N_{MQ}
\ .
\end{equation}
Since $n\ge 1$, the above implies the condition
\begin{equation}
\label{eq:GS-positive-condition}
\frac{\mu\, M}{\mpl^{2}}
-
\frac{1}{2}\,
\sqrt{1+ \frac{4\,Q^{2}\,\mu^{2}}{\mpl^{3}\,\lp}}
\ge
\frac{1}{2}
\ ,
\end{equation}
that is, the ADM mass and charge must satisfy~\footnote{For $Q=0$, this condition
implies that the Schwarzschild radius must be sufficiently larger than the Compton length
of one dust particle, $\Rh=2\,\gn\,M>2\,\hbar/\mu$ (see also
Refs.~\cite{Casadio:2013tma,Casadio:2013aua,Calmet:2015pea}).}
\begin{equation}
\label{eq:m2-constraint}
\gn\,M^{2}
\ge
Q^{2}
+
M\left({\hbar}/{\mu}\right)
>
Q^{2}
\ ,
\end{equation}
which is therefore stronger than the classical condition~\eqref{condR+-} for the existence of
(two) horizons.
In particular, Eq.~\eqref{eq:m2-constraint} excludes the exact classical extremal
case $R_-=R_+$ obtained for $\gn\,M^2=Q^2$ (see Appendix~\ref{A:condQ} for more details).
\par
\begin{figure}[t]
\begin{center}
\includegraphics[width=0.48\textwidth]{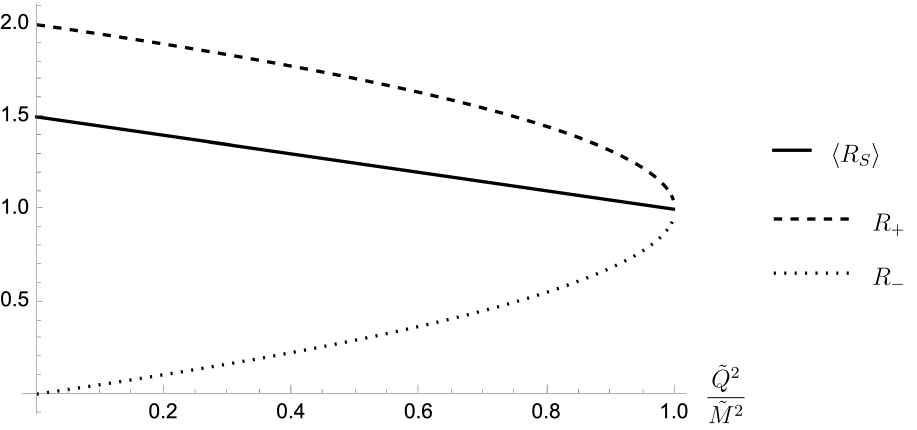}
$\ $
\includegraphics[width=0.48\textwidth]{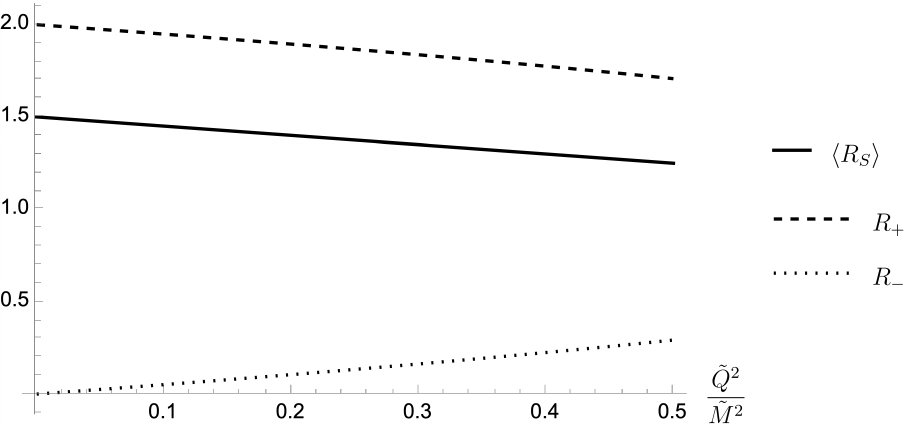}
\end{center}
\caption{\tb Ball radius $\expec{\hat R_{\rm s}}$ and horizon radii $R_\pm$ for $\tilde M=10^3$
with $\tilde\mu=1$ (left panel) and $\tilde\mu=2\cdot 10^{-3}$ (right panel):
$R_-<\expec{\hat R_{\rm s}}<R_+$ for all allowed values of the charge $Q$
that satisfy the condition~\eqref{eq:m2-constraint} depending on $\tilde\mu$.
(Tilded quantities are dimensionless variables defined in Footnote~\ref{ftn6}.)}
\label{fig:1a}
\end{figure}

{\tb The expectation value of the size of the ball in the ground state is obtained by direct
substitution of Eqs.~\eqref{eq:n-gs} and~\eqref{g_mu} into Eq.~\eqref{eq:expct-r}, and yields}
\begin{equation}
\label{eq:expectR-singleLayer}
\expec{\hat R_{\rm s}}
\equiv
\bra{N_{MQ}}\hat R_{\rm s}\ket{N_{MQ}}
=
\frac{3}{2}\,\gn\,M
\left(1
-
\frac{Q^{2}}{3\,\gn\,M^2}
\right)
\ ,
\end{equation}
so that the core has a finite radius of size
\be
\gn\,M<\expec{\hat R_{\rm s}}\le \frac{3}{2}\,\gn\,M
\ .
\ee
{\tb On comparing Eqs.~\eqref{eq:expectR-singleLayer} and~\eqref{eq:RN-hor} we see that
the ball} in the ground state lies inside the classical outer horizon if $\expec{\hat R_{\rm s}}<R_+$.
Given the condition in Eq.~\eqref{eq:m2-constraint}, this implies
\be
\sqrt{1
-
\frac{Q^{2}}{\gn\,M^2}}
<
2
\ ,
\ee
which is always satisfied and we can say that all consistent ground states represent black holes.
It is then interesting to investigate if the ground state can lie inside the classical
inner horizon, $\expec{\hat R_{\rm s}}<R_-$.
This would happen for
\be
\sqrt{
1-
\frac{Q^{2}}{\gn\,M^2}
}
<
-2
\ ,
\ee
which cannot be met, again because of the condition {\tb given by Eq.~\eqref{eq:m2-constraint}}.
The overall conclusion is that all consistent quantum ground states are black
holes without a Cauchy inner horizon (see Fig.~\ref{fig:1a} and further comments
below~\footnote{\tb In numerical plots we employ dimensionless quantities given by $\tilde M=M/\mpl$,
$\tilde \mu=\mu/\mpl$, $\tilde Q^2=Q^2/\lp\,\mpl$, $\expec{\tilde R_{\rm s}} = \expec{\hat R_{\rm s}}/\lp$.
\label{ftn6}}).
\begin{figure}[t]
\begin{center}
\includegraphics[scale=0.8]{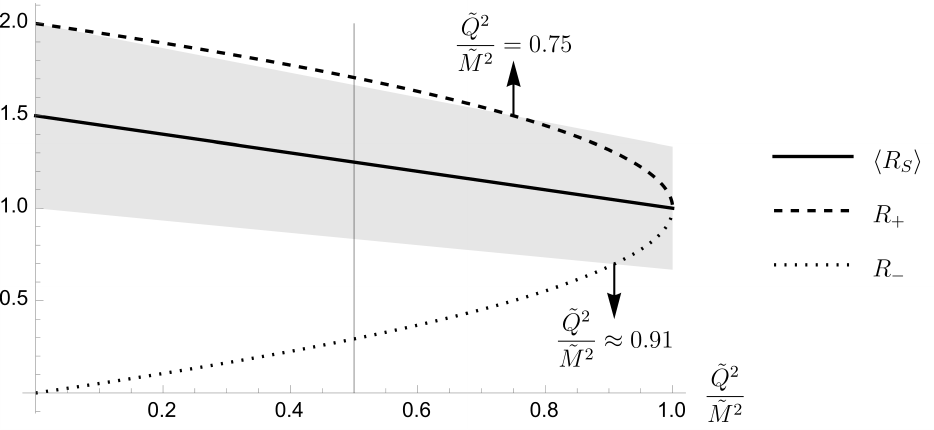}
\end{center}
\caption{
\tb Ball radius $\expec{\hat R_{\rm s}}$ and horizon radii $R_\pm$ as functions of the charge-to-mass ratio.
Note that $R_-<\expec{\hat R_{\rm s}}<R_+$ for all classically allowed values of the charge $\tilde Q^2\le \tilde M^2$.
Shaded band represents the fuzzy region bounded by $\expec{\hat R_{\rm s}}\pm\Delta R_{\rm s}$.
Arrows indicate values where the (event or Cauchy) horizon radius enters the fuzzy region.
Thin vertical line indicates the maximum value of the charge-to-mass ratio from the right panel of
Fig.~\ref{fig:1a}.
(Tilded quantities are dimensionless variables defined in Footnote~\ref{ftn6}.)
}
\label{fig:1b}
\end{figure}
\par
{\tb We conclude this part by looking at the uncertainty~\eqref{eq:r-var} for the size of the
the ground state with $n=N_{MQ}$ for astrophysical objects with
\begin{equation}
\label{eq:beta-astrph}
\beta_M=\mu\, M/\mpl^2\gg 1
\ .
\end{equation}
In this case, we expect that the charge is much smaller than the mass,
that is $Q^2\ll \gn\,M^2$, so that $\alpha\sim 1$ and $N_{MQ}\gg 1$.
We can expand Eq.~\eqref{eq:r-var} in this regime to obtain
\be
\frac{\Delta R_{\rm s}}{\expec{\hat R_{\rm s}}}
\simeq
\frac{1}{3}
+
\frac{11+\alpha^{2}}{36\,\beta_{M}^{2}}
\simeq
\frac{1}{3}
\left[
1
+
\frac{\mpl^4}{\mu^2\,M^2}
\left(
1+
\frac{2 \mu^{2}\,Q^{2}}{3\,\lp\,\mpl^{3}}
\right)
\right]
\simeq \frac{1}{3}
\ ,
\label{dR}
\ee
where the last approximation comes from our assumption in Eq.~\eqref{eq:beta-astrph}.
In particular, we notice that the size $R_{\rm s}$ of the quantum ball is fuzzy and should
approximately lie within the interval $|R_{\rm s}-\expec{\hat R_{\rm s}}|\le \Delta R_{\rm s}$,
where
\be
\expec{\hat R_{\rm s}}
-
\Delta R_{\rm s}
\simeq
\frac{2}{3}\,\expec{\hat R_{\rm s}}
\simeq
\gn\,M
-
\frac{Q^{2}}{3\,M}
\label{r-dR}
\ee
and
\be
\expec{\hat R_{\rm s}}
+
\Delta R_{\rm s}
\simeq
\frac{4}{3}\,\expec{\hat R_{\rm s}}
\simeq
2\,\gn\,M
-
\frac{2\,Q^{2}}{3\,M}
\ .
\label{r+dR}
\ee
This range is represented by the shaded region in Fig.~\ref{fig:1b} (for the same values of
$M$ and $\mu$ used in the left panel of Fig.~\ref{fig:1a} so that the maximum value
of $Q^2\simeq \gn\,M^2$).
\par
We notice in particular that the upper boundary~\eqref{r+dR} is shorter than the classical event
horizon at $r=R_+$ if
\be
1
-
\frac{2\,Q^{2}}{3\,\gn\,M^2}
<
\sqrt{1-\frac{Q^2}{\gn\,M^2}}
\ ,
\ee
equivalent to $4\,Q^2 < 3\,\gn\,M^2$.
The matter core could therefore be found with a larger radius than the outer horizon $R_+$ for
${Q^{2}}/{\gn\,M^{2}}\gtrsim 0.75$ (see the upward arrow in Fig.~\ref{fig:1b}).
This also implies that a dust ball of charge close to the extremal classical case $Q^2=\gn\,M^2$
would very likely not be a classical black hole at all,~\footnote{Similar results were previously obtained
in Refs.~\cite{Casadio:2022ndh,Casadio:2015rwa}.}
meaning that the probability of finding dust particles placed at $r>R_+$ is not negligible
(see, {\em e.g.}~Refs.~\cite{Casadio:2013tma,Casadio:2013aua,Feng:2024nvv}).
Likewise, the lower boundary~\eqref{r-dR} is shorter than the inner horizon $R_-$ if
\be
\sqrt{1-\frac{Q^2}{\gn\,M^2}}
<
\frac{Q^2}{3\,\gn\,M^2}
\ ,
\ee
which means that dust particles at the surface of the matter core could be found inside the classical
Cauchy horizon $R_-$ with significant probability provided ${Q^{2}}/{\gn\,M^{2}} \gtrsim 0.91$
(see the downward arrow in Fig.~\ref{fig:1b}).
The relevance of these results is that they further clarify the proper quantum nature of the matter
core and ensuing geometry.}
\section{Multilayer core}
\label{sec:mlayer_core}
\setcounter{equation}{0}
We now want to improve on the previous description of the dust ball by considering
several layers surrounding the innermost core where the charge is located,
as described in the Introduction.
The solution for Eq.~\eqref{eq:geodesic} is of the same form as
Eq.~\eqref{eq:wavefunction-sol}, namely
\begin{equation}
\label{eq:wavefunction-mlayer}
\pro{R_{i}}{n_i\alpha}
\equiv
\Psi_{n_{i}\alpha}(R_{i})
=
A_{n_i\alpha}\, e^{-\frac{\gamma_{i}}{2}\,R_{i}}\,
R_{i}^{\frac{\alpha-1}{2}}\,
L_{n_{i}-1}^{\alpha}(\gamma_{i}\,R_{i})
\ .
\end{equation}
{\tb where $r=R_{i}$ is the areal radius of dust particles located on the surface that identifies the
inner border of the $i^{\rm th}$ layer and the integer $n_{i}\ge 1$ labels their quantum state.
Similarly to what was done in Section~\ref{sec:QRN_core}, we define the
parameters}
\be
\label{eq:nint-mlayer}
\beta_{n_i\alpha}
\equiv
n_{i}+\frac{\alpha-1}{2}
=
\frac{2\,\mu^{2}\, M_{i}}{\gamma_{i}\, \lp\, \mpl^{3}}
\ ,
\ee
{\tb  where now $ \gamma_{i} $ is associated with the energy levels of each shell $ \varepsilon_{i} $
by}
\begin{equation}
\label{eq:gamma-mlayer}
\gamma_{i}^{2}
=
-\frac{8\,\mu\,\varepsilon_{i}}{\mpl^{2}\,\lp^{2}}
\ ,
\end{equation}
and $\alpha$ is still given by Eq.~\eqref{eq:k2-Q} for all of the layers.
\par
Substituting everything back in Eq.~\eqref{eq:wavefunction-mlayer} we obtain the spectrum
\begin{equation}
\Psi_{n_{i}\alpha}(R_{i})
=
A_{n_i\alpha}\,
\exp\!{\left(-\frac{\mu^{2}\,M_{i}\,R_i}{\beta_{n_i\alpha}\,\mpl^{3}\,\lp}\right)}\,
R_{i}^{\frac{\alpha-1}{2}}\,
L_{n_{i}-1}^{\alpha}\!\left(\frac{2\,\mu^{2}\,M_{i}\,R_{i}}{\beta_{n_i\alpha}\,\mpl^{3}\,\lp}\right)
\ ,
\label{psi_i}
\end{equation}
where
\be
A_{n_i\alpha}^2
=
\frac{\Gamma(n_{i})}{8\,\beta_{n_i\alpha}\,\pi^{2}\,\Gamma^3(\alpha+n_{i})}
\left( \frac{2\, \mu^{2}\,M_{i}}{\beta_{n_i\alpha}\,\mpl^{3}\,\lp}\right)^{\alpha+2}
\ ,
\ee
with eigenvalues
\begin{equation}
\varepsilon_{n_i\alpha}
=
-\frac{ \mu^{3}\,M_i^{2} }{2\,\beta_{n_i\alpha}^{2}\, \mpl^{4}}
\ .
\end{equation}
{\tb These results are a simple generalisation of the approach to the global radius of the
ball employed in Section~\ref{sec:QRN_core}, and will be used it to describe a compact core for the black hole
formed by layers identified by particles located on their inner borders.}
\subsection{Single ground states}
{\tb Following the procedure used in Section~\ref{ssec:gs-wavefunction}, particles in the ground state
of each layer will have energy $\epsilon_i=-\mu/2$, equivalent to $E_i^2=0$, from Eq.~\eqref{eq:geodesic_gen}.}
Correspondingly, we obtain $\gamma_i=\gamma_\mu$ in Eq.~\eqref{g_mu} and
\be
\beta_{n_i\alpha}
=
\frac{\mu\,M_i}{\mpl^2}
\equiv
\beta_{i}
\ .
\ee
The quantum number for dust particles in the ground state of each layer is therefore
given by Eq.~\eqref{eq:n-integer} with $M=M_i$ and reads
\begin{equation}
\label{eq:n_i}
n_i
=
\frac{\mu\, M_i}{\mpl^{2}}
+
\frac{1}{2}
\left(
1-
\sqrt{1+ \frac{4\,Q^{2}\,\mu^{2}}{\mpl^{3}\,\lp}}
\right)
\equiv
N_{i}
\ .
\end{equation}
From this result, we {\tb use Eq.~\eqref{eq:expct-r} to obtain the expectation value of each shell},
\be
\expec{\hat R_i}
\simeq
\frac{3}{2}\,\gn\,M_i
\left(1
-
\frac{Q^2}{3\,\gn\,M_i}
\right)
\ ,
\label{Ri}
\ee
and, from the leading term in Eq.~\eqref{dR} for $\beta_i\gg 1$,
the uncertainty
\be
\Delta R_i=\sqrt{\expec{\hat R_i^2}-\expec{\hat R_i}^2}
\simeq
\frac{1}{3}\,
\expec{\hat R_i}
\ .
\label{dRi}
\ee
These ground states are well defined provided the charge and discrete mass function
$M_i$ satisfy the condition {\tb given by Eq.~\eqref{eq:m2-constraint}}, that is
\be
\label{eq:m2-constraintL}
\gn\,M_i^{2}
\ge
Q^{2}
+
M_i\left({\hbar}/{\mu}\right)
>
Q^{2}
\ .
\ee
Since $M_i> M_{i-1}$, the strongest constraint comes from $i=1$, that is
\be
Q^2
<
\gn\,\mu_0^2
\ ,
\label{sQc}
\ee
where we used $M_1=\mu_0$ for the ADM mass of the central core.
\subsection{Global ground state}
\label{ss:globalGS}
{\tb Having generalised the construction for the global radius and the ground state expressions
to individual layers, we can now study a matter core of dust for the charged black holes
made of nested layers around a central charged core.}
In particular, the number $N$ of layers and the (discrete) mass function $m(R_i)=M_i$ for the
ground state is still to be determined.
Like in Ref.~\cite{Casadio:2023ymt}, we can find a self-consistent ground state
by requiring that each layer of radius $\expec{\hat R_i}$ in Eq.~\eqref{Ri} have
a thickness given by the uncertainty $\Delta R_i$ in Eq.~\eqref{dRi}.
\par
The above assumption implies that
\begin{equation}
\expec{\hat R_{i+1}}
\simeq
\expec{\hat R_i}
+
\Delta R_{i}
\simeq
\frac{4}{3}\,
\expec{\hat R_i}
\ ,
\end{equation}
which reproduces Eq.~\eqref{eq:expectR-singleLayer} for the
surface of the ball with $i=N$.
This straightforwardly leads to the condition
\begin{equation}
\label{eq:mi-mi1}
3\,M_{i+1}
-
\frac{Q^{2}}{\gn\,M_{i+1}}
\simeq
4\,M_{i}
-
\frac{4\,Q^{2}}{3\,\gn\,M_{i}}
\ ,
\end{equation}
which indeed reduces to the expected relation for the neutral case $3\,M_{i+1}=4\,M_i$
for $Q=0$~\cite{Casadio:2023ymt}.
\par
We will solve Eq.~\eqref{eq:mi-mi1} for the discrete mass function numerically.
{\tb It is again convenient to use the dimensionless quantities defined in Footnote~\ref{ftn6},
so that the equation now reads}
\begin{equation}
\label{eq:recur_mass}
3\,\tilde{M}_{i+1}
-
\frac{\tilde{Q}^{2}}{\tilde{M}_{i+1}}
-4\,\tilde{M}_{i}
+
\frac{4\,\tilde{Q}^{2}}{3\,\tilde{M}_{i}}
=
0
\ .
\end{equation}
The constraints~\eqref{eq:m2-constraintL} also read
\begin{equation}
\label{eq:Q_constraint}
\tilde{M}_{i}^{2}
-\frac{\tilde M}{\tilde{\mu}}
-
\tilde{Q}^{2}
>0
\ .
\end{equation}
We can solve Eqs.~\eqref{eq:recur_mass} recursively by starting from the total ADM mass
$M = M_{N+1}$.
The mass $M_N$ inside the outermost layer is then given by the positive solution
of Eq.~\eqref{eq:recur_mass} with $i=N$, and the process is repeated by solving
Eq.~\eqref{eq:recur_mass} with $M_{N+1}\mapsto M_{N}$, and so on for $i\mapsto i-1$.
In this way, the mass inside the $i^{\rm th}$ layer can be obtained as long as
Eq.~\eqref{eq:Q_constraint} is satisfied, say up to $i=N-k$, for which we stop and assume
$M_{N-k}=M_0$ is the central core containing the charge $Q$.
The number of layers $N=k$ for a given ADM mass and charge is thus determined consistently.
\par
\begin{figure}[ht]
\begin{center}
\includegraphics[width=0.45\textwidth]{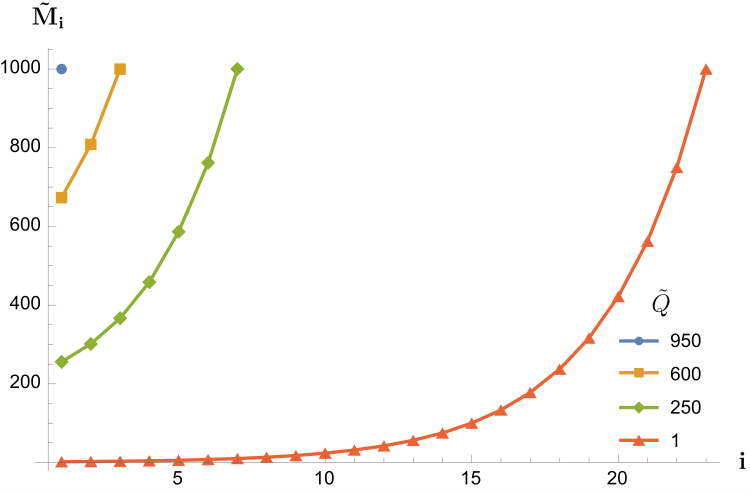}
$\ $
\includegraphics[width=0.45\textwidth]{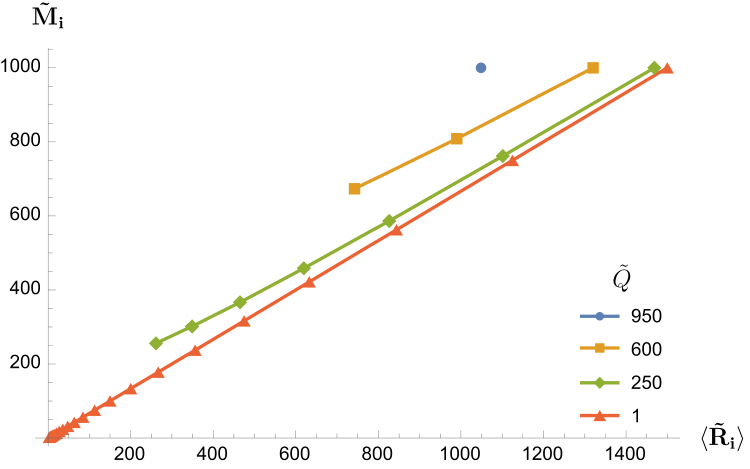}
\end{center}
\caption{Discrete mass function $m=M_i$ as function of the layer $i$ (left panel)
and areal radius $\expec{\tilde R_i}$ (right panel) for $\tilde{M} = 10^3$, $\tilde{\mu} = 1$
and different values of $ \tilde{Q}$.}
\label{fig:plot2}
\end{figure}
\begin{figure}[ht]
\begin{center}
\includegraphics[width=0.45\textwidth]{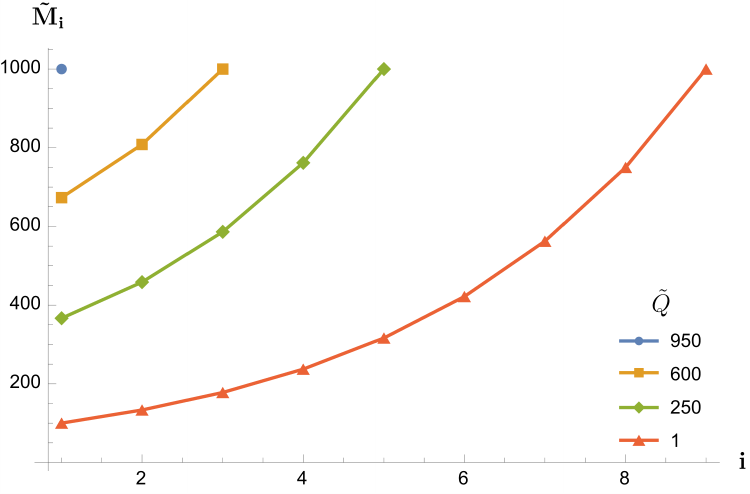}
$\ $
\includegraphics[width=0.45\textwidth]{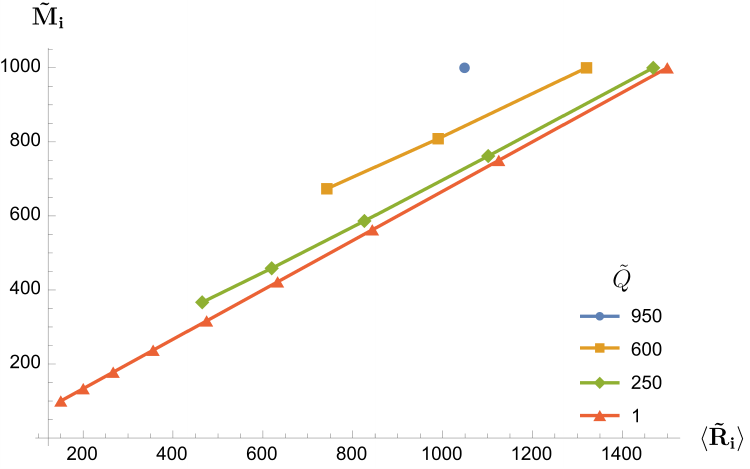}
\end{center}
\caption{Discrete mass function $m=M_i$  as function of the layer $i$ (left panel)
and areal radius $\expec{\tilde R_i}$ (right panel) for $\tilde{M} = 10^3$, $\tilde{\mu} = 10^{-2}$
and different values of $ \tilde{Q}$.}
\label{fig:plot3}
\end{figure}

{\tb  A few results of the above procedure are displayed in Figs.~\ref{fig:plot2} and~\ref{fig:plot3},
for different values of $\tilde M$, $\tilde\mu$ and $\tilde Q$ chosen so that $M\gg\mu$ and
the plots are easier to read.
By comparing the different plots, one can see that heavier dust particles lead to the formation
of more layers, and most of these layers are closer to the central core [in agreement with
	the uncertainty~\eqref{dRi}].}
As expected, the number of layers decreases for larger charge $\tilde Q$, since the
inequality~\eqref{eq:Q_constraint} saturates for smaller values of $k$.
It is particularly interesting to notice that the (dimensionless) discrete mass function
is essentially linear in the (dimensionless) layer radius
\begin{equation}
\label{eq:adimR}
\expec{\tilde{R}_{i}}
=
\frac{3}{2}\,\tilde{M}_{i}
\left(1
-
\frac{\tilde{Q}^{2}}{3\,\tilde{M}_{i}^2}
\right)
\ .
\end{equation}
This feature also appears in the neutral case~\cite{Casadio:2023ymt}.
\par
{\tb The right panels in Figs.~\ref{fig:plot2} and~\ref{fig:plot3} show the discrete mass function $M=M(R_i)$,
and we can clearly see a linear behaviour, except near the extremal case $\tilde Q\simeq \tilde M$.
In that limiting regime, our analysis implies that the core would only admit one layer (on top of the central charged core),
which is also in qualitative agreement with the fuzziness of the ball radius discussed in
Section~\ref{ssec:gs-wavefunction}.}
\subsection{Effective metric and energy-momentum tensor}
The discrete mass function $M=M(R_i)$ obtained numerically in the previous section can be approximated
by the continuous function
\be
m
\simeq
K\,r
\ ,
\ee
where $K\simeq 2/3\,\gn$ for $Q^2\ll \gn\,M^2$ (see Figs.~\ref{fig:plot2} and~\ref{fig:plot3}).
This mass function corresponds to an effective quantum metric inside the dust ball given by Eq.~\eqref{eq:rn}
with~\footnote{The metric signature is $+-++$ ($r$ is a time coordinate) for $r<\expec{\hat R_{\rm s}}<R_+$.}
\be
f
=
f_{\rm q}
\simeq
-1+2\,\gn\,K-\frac{\gn\,Q^2}{r^2}
\ .
\label{fqQ}
\ee
The positive zeros of $f_{\rm q}$ would locate the possible horizons inside the dust ball,
namely
\be
r_-^2
\simeq
\frac{\gn\,Q^2}{3}
\ .
\ee
However, the charge $Q$ must satisfy the condition~\eqref{sQc}, which implies that $r_-^2<{\gn^2\,\mu_0^2}/{3}$
and lies inside the central core of mass $M_1=\mu_0$, where the effective metric~\eqref{fqQ} cannot be employed.
We also remark that the geometry in the region outside the ball will be given by the vacuum Reissner-Nordstr\"om
metric~\eqref{fQ} for $r>\expec{\hat R_{\rm s}}$ and one still has the outer event horizon $R_+$.~\footnote{The
precise matching of $f_{\rm q}$ with $f_{\rm RN}$ at $r=\expec{\hat R_{\rm s}}$ requires a better approximation
for both effective metrics which is left for future developments.}
These results are in agreement with the analysis in Sections~\ref{ssec:expect-values} and~\ref{ss:globalGS}.
\par
From the above metric, we can then compute the Einstein tensor $G^\mu_{\ \nu}=8\,\pi\,\gn\,T^{\mu}_{\ \nu}$
and determine the effective energy density and pressure
\be
\rho
=
-T^r_{\ r}
\simeq
\frac{K}{4\,\pi\,r^{2}}
+
\frac{Q^{2}}{8\,\pi\,r^{4}}
=
-T^t_{\ t}
\simeq
-p_r
\ ,
\label{rho}
\ee
and the effective tension
\be
p_\perp
=
T^\theta_{\ \theta}
\simeq
\frac{Q^{2}}{8\,\pi\,r^{4}}
\ .
\label{pt}
\ee
These effective quantities reproduce the expressions in Ref.~\cite{Casadio:2023ymt} for the neutral case $Q=0$
with the same value of $K\simeq 2/3\,\gn$.~\footnote{From Figs.~\ref{fig:plot2} and~\ref{fig:plot3},
we see that $K$ decreases slightly for larger $Q$.}
\par
It is important to notice that the term proportional to $Q^2$ in the energy density {\tb Eq.~\eqref{rho} }
is not integrable but would correspond to the electric field contribution to the standard Reissner-Nordstr\"om
singularity if the charge were localised at $r=0$.
As we discussed in the Introduction, we instead assume that the charge is distributed over the
innermost core of finite radius $\expec{\hat R_1}\simeq \gn\,M_1>0$ given by Eq.~\eqref{Ri}
with $i=1$.
Eqs.~\eqref{rho} and \eqref{pt} therefore only hold for $\expec{\hat R_1}<r<\expec{\hat R_{\rm s}}$
(see Figs.~\ref{fig:plot2} and~\ref{fig:plot3} for examples of values taken by $\expec{\hat R_1}$).
\section{Concluding remarks}
\label{sec:conc}
\setcounter{equation}{0}
We investigated a dust ball of mass $M$ with electric charge $Q$ localised inside a central massive core
surrounded by electrically neutral layers of dust particles governed by the general relativistic dynamics.
All dust particles would classically fall along radial geodesics in a Reissner-Nordstr\"om
metric with suitable mass function, which results in the Hamiltonian constraint {\tb Eq.}~\eqref{eq:geodesic}.
\par
This constraint equation is canonically quantised similarly to the one describing the motion
of the electron in the hydrogen atom and yields the discrete spectrum {\tb Eq.}~\eqref{psi_i} for the dust particles
in terms of generalised associated Laguerre polynomials with indices determined by the mass
(function) and electric charge.
The thickness of each layer is also related to the quantum uncertainty in the (radial) localisation
of particles therein.
\par
The system was solved explicitly for the ground state, which can exist only if the charge
and mass function satisfy the bound in Eq.~\eqref{eq:m2-constraintL}.
In particular, these conditions ensure that the dust ball in the ground state is a black hole,
since all of the dust is localised inside the outer horizon of the Reissner-Nordstr\"om metric,
but tidal forces remain finite everywhere, so that there is no physically dangerous singularity
in the centre.
Moreover, the classically extremal case cannot be realised and the presence of an inner Cauchy
horizon is further excluded.
These two results are particularly noteworthy and stem from the finite spatial size of the
ground state.
\par
The main difference with respect to the neutral case studied in
Refs.~\cite{Casadio:2021cbv,Casadio:2023ymt} is that the quantum numbers for the
layers in the ground state are always very large for a ball of astrophysical mass
if $Q=0$ but can be of order one if the charge approaches the limiting value allowed
by the bound {\tb Eq.}~\eqref{eq:m2-constraint}.
Beside this technical aspect, the addition of an electric charge $Q$ at the centre does not alter
the main features previously found in the completely neutral model, which further supports the
physical solidity of our approach to quantisation.
\par
If the ground states obtained in Ref.~\cite{Casadio:2023ymt} and in Section~\ref{ss:globalGS}
were indeed to describe the endpoint of the gravitational collapse, one could conclude that no
physically dangerous singularity is formed and the absence of a Cauchy horizon further removes
known potential sources of instability for those configurations.
The existence of a very large (quantum) matter core inside the horizon can then have potentially
very interesting consequences for physical processes occurring in the outer region that can
therefore be detected, as shown, {\em e.g.}~in Ref.~\cite{Arrechea:2024nlp}.
{\tb The detailed analysis of such effects is left for future publications.}
\subsection*{Acknowledgments}
R.C.~is partially supported by the INFN grant FLAG.
R.dR.~is grateful to the S\~ao Paulo Research Foundation FAPESP (grants No.~2024/05676-7
and No.~2021/01089-1) and CNPq (grants No.~303742/2023-2 and No.~401567/2023-0)
for partial financial support.
A.G.~is supported in part by the Science and Technology Facilities Council (grants
No.~ST/T006048/1 and ST/Y004418/1).
P.M.~thanks FAPESP for the financial support (grants No.~2022/12401-9 and No.~2023/12826-2).
The work of R.C.~and A.G.~has also been carried out in the framework of activities of the
National Group of Mathematical Physics (GNFM, INdAM).
\appendix
\section{Charge constraints for the ground state and extremal case}
\label{A:condQ}
\setcounter{equation}{0}
The expectation value $\expec{\hat R_{\rm s}}$ of the ball radius in Eq.~\eqref{eq:expectR-singleLayer}
is positive provided
\be
12\,\beta_{M}^{2}-\alpha^{2}+1
>
0
\ .
\label{eq:finite_var}
\ee
From Eq.~\eqref{eq:r-var} with $\beta_{n\alpha}=\beta_M=\mu\,M/\mpl^2$,
its uncertainty $\Delta R_{\rm s}$ is also positive if
\be
16\,\beta_{M}^{2}\left(\beta_{M}^{2}+2\right)-\left(\alpha^{2}-1\right)^{2}
>0
\ .
\label{eq:real_var}
\ee
Using the definition {\tb in Eq.}~\eqref{eq:k2-Q}, these inequalities read
\begin{equation}
\label{eq:realVarQ}
\frac{Q^{2}}{\gn\,M^{2}}
<
\sqrt{1+\frac{2\,\mpl^4}{\mu^{2}\,M^{2}}}
\end{equation}
and
\begin{equation}
\label{eq:finiteVarQ}
Q^{2}
<
3\,\gn\,M^{2}
\ .
\end{equation}
The conditions in Eqs.~\eqref{eq:realVarQ} and~\eqref{eq:finiteVarQ} are satisfied
if Eq.~\eqref{eq:m2-constraint} holds.
Similar results are obtained for individual layers.
\par
We next notice that the condition for the existence of the quantum spectrum in
Eq.~\eqref{eq:m2-constraint} implies that
\beq
\label{bound1}
1
\le
\alpha
\le
2\,\beta_M-1
\ .
\eeq
We already considered the more astrophysical relevant case $\alpha\sim 1$
for the dust ball with a relatively small amount of charge in Section~\ref{ssec:gs-wavefunction}.
Here, we shall instead study the ground state in the opposite case
when $\alpha\sim \beta_M$, with $N_{MQ}\sim 1$.
From $\beta_M=\mu\,M/\mpl^2$, we have
\begin{equation}
\alpha
\sim
\frac{\mu\,M}{\mpl^{2}}
\end{equation}
and Eq.~\eqref{eq:k2-Q} then implies
\begin{equation}
\label{eq:ext_Q}
Q^2
\sim
\gn\,M^2
\ .
\end{equation}
which approaches the extremal case (see Fig.~\ref{fig:1a}).
For the same limiting case one obtains
\begin{equation}
\lim_{n\to 1}\,
\frac{\Delta R_{\rm s}}{\bra{n\alpha}\hat R_{\rm s}\ket{n\alpha}}
=
\frac{\sqrt{4\,\mu^2\,M^2/\mpl^4+7+\mpl^2/\mu\,M}}{2\left(1+2\,\mu\,M/\mpl^2\right)}
\simeq
\frac{1}{2}
\ ,
\end{equation}
where we used $\beta_M\gg 1$.
We can in general conclude that the uncertainty
\be
\frac{1}{3}
<
\frac{\Delta R_{\rm s}}{\bra{n\alpha}\hat R_{\rm s}\ket{n\alpha}}
<
\frac{1}{2}
\ ,
\ee
for $\mu\,M/\mpl^2\gg 1$.
\bibliographystyle{unsrt}

\begin{thebibliography}{99}
%
%
\bibitem{ADM}
R.~L.~Arnowitt, S.~Deser and C.~W.~Misner,
Phys.\ Rev.\  {\bf 116} (1959) 1322.
%
\bibitem{Wald:1984rg}
R.~M.~Wald,
``General Relativity,''
(Chicago University Press, 1984)
%
\bibitem{HE}
S.~W.~Hawking and G.~F.~R.~Ellis,
``The Large Scale Structure of Space-Time,''
(Cambridge University Press, Cambridge, 1973)
%
\bibitem{geroch}
R.~P.~Geroch and J.~H.~Traschen,
Phys. Rev. D \textbf{36} (1987) 1017.
%
\bibitem{balasin}
H.~Balasin and H.~Nachbagauer,
Class. Quant. Grav. \textbf{10} (1993) 2271
[arXiv:gr-qc/9305009 [gr-qc]].
%
\bibitem{McNamara}
J.~M.~McNamara,
Proc. R. Soc. London \textbf{A358} (1978) 449.
%
\bibitem{Gursel:1979zza}
Y.~Gursel, V.~D.~Sandberg, I.~D.~Novikov and A.~A.~Starobinsky,
Phys. Rev. D \textbf{19} (1979) 413.
%
\bibitem{chandra}
S.~Chandrasekhar and J.~B.~Hartle,
Proc. R. Soc. London \textbf{A284} (1982) 301.
%
\bibitem{Poisson:1989zz}
E.~Poisson and W.~Israel,
Phys. Rev. Lett. \textbf{63} (1989) 1663.
%
\bibitem{Gnedin:1993nau}
M.~L.~Gnedin and N.~Y.~Gnedin,
Class. Quant. Grav. \textbf{10} (1993) 1083.
%
\bibitem{Brady:1995ni}
P.~R.~Brady and J.~D.~Smith,
Phys. Rev. Lett. \textbf{75} (1995) 1256
[arXiv:gr-qc/9506067 [gr-qc]].%
%
\bibitem{Burko:1997zy}
L.~M.~Burko,
Phys. Rev. Lett. \textbf{79} (1997) 4958
[arXiv:gr-qc/9710112 [gr-qc]].
%
\bibitem{Marolf:2011dj}
D.~Marolf and A.~Ori,
Phys. Rev. D \textbf{86} (2012) 124026
[arXiv:1109.5139 [gr-qc]].
%
\bibitem{Eilon:2016osg}
E.~Eilon and A.~Ori,
Phys. Rev. D \textbf{94} (2016) 104060
[arXiv:1610.04355 [gr-qc]].
%
\bibitem{Carballo-Rubio:2024dca}
R.~Carballo-Rubio, F.~Di Filippo, S.~Liberati and M.~Visser,
[arXiv:2402.14913 [gr-qc]].
%
\bibitem{Balbinot:2023grl}
R.~Balbinot and A.~Fabbri,
Phys. Rev. D \textbf{108} (2023) 045004
[arXiv:2303.11039 [gr-qc]].
%
\bibitem{Casadio:2019tfz}
R.~Casadio and A.~Giusti,
Phys. Lett. B \textbf{797} (2019) 134915
[arXiv:1904.12663 [gr-qc]].
%
\bibitem{Casadio:1998yr}
R.~Casadio,
Int. J. Mod. Phys. D \textbf{9} (2000) 511
[arXiv:gr-qc/9810073 [gr-qc]].
%
\bibitem{Kuntz:2019lzq}
I.~Kuntz and R.~Casadio,
Phys. Lett. B \textbf{802} (2020), 135219
[arXiv:1911.05037 [hep-th]].
%
\bibitem{Kuntz:2019gup}
I.~Kuntz and R.~da Rocha,
Eur. Phys. J. C \textbf{79} (2019) 447
[arXiv:1903.10642 [hep-th]].
%
\bibitem{Haggard:2014rza}
H.~M.~Haggard and C.~Rovelli,
Phys. Rev. D \textbf{92} (2015) 104020
[arXiv:1407.0989 [gr-qc]].
%
\bibitem{Bonanno:2023rzk}
A.~Bonanno, D.~Malafarina and A.~Panassiti,
Phys. Rev. Lett. \textbf{132} (2024) 031401
[arXiv:2308.10890 [gr-qc]].
%
\bibitem{Lemaitre:1927zz}
G.~Lemaitre,
Annales Soc. Sci. Bruxelles A \textbf{47} (1927) 49.
%
\bibitem{Tolman:1934za}
R.~C.~Tolman,
Proc. Nat. Acad. Sci. \textbf{20} (1934) 169.
%
\bibitem{Bondi:1947fta}
H.~Bondi,
Mon. Not. Roy. Astron. Soc. \textbf{107} (1947) 410.
%
\bibitem{Oppenheimer:1939ue}
J.~R.~Oppenheimer and H.~Snyder,
Phys. Rev. \textbf{56} (1939) 455.
%
\bibitem{EIH}
A.~Einstein, J.~Grommer,
\textit{Allgemeine Relativit\"{a}tstheorie und Bewegungsgesetz},
Sitzber. Preuss. Akad. Wiss. Berlin, {\bf 2} (1927);
A.~Einstein, L.~Infeld, B.~Hoffmann,
\textit{The Gravitational Equations and the Problem of Motion},
Ann. Math., {\bf 39}, 65 (1938).
%
\bibitem{Kiefer:2004xyv}
C.~Kiefer,
``Quantum gravity''
(Clarendon Press, 2004)
%
\bibitem{Casadio:1998ta}
R.~Casadio,
Phys. Rev. D \textbf{58} (1998) 064013
[arXiv:gr-qc/9804021 [gr-qc]].
%
\bibitem{Vaz:2011zz}
C.~Vaz and L.~Witten,
Gen. Rel. Grav. \textbf{43} (2011) 3429
[arXiv:1111.6821 [gr-qc]].
%
\bibitem{Kiefer:2019csi}
C.~Kiefer and T.~Schmitz,
Phys. Rev. D \textbf{99} (2019) 126010
[arXiv:1904.13220 [gr-qc]].
%
\bibitem{Piechocki:2020bfo}
W.~Piechocki and T.~Schmitz,
Phys. Rev. D \textbf{102} (2020) 046004
[arXiv:2004.02939 [gr-qc]].
%
\bibitem{Schmitz:2020vdr}
T.~Schmitz,
Phys. Rev. D \textbf{103} (2021) 064074
[arXiv:2012.04383 [gr-qc]].
%
\bibitem{Husain:2022gwp}
V.~Husain, J.~G.~Kelly, R.~Santacruz and E.~Wilson-Ewing,
Phys. Rev. D \textbf{106} (2022) 024014
[arXiv:2203.04238 [gr-qc]].
%
\bibitem{Giesel:2022rxi}
K.~Giesel, M.~Han, B.~F.~Li, H.~Liu and P.~Singh,
Phys. Rev. D \textbf{107} (2023) 044047
[arXiv:2212.01930 [gr-qc]].
%
\bibitem{Jacobson:1995ab}
T.~Jacobson,
Phys. Rev. Lett. \textbf{75} (1995) 1260
[arXiv:gr-qc/9504004 [gr-qc]].
%
\bibitem{Bardeen:1973gs}
J.~M.~Bardeen, B.~Carter and S.~W.~Hawking,
Commun. Math. Phys. \textbf{31} (1973) 161.
%
\bibitem{bekenstein}
J.D.~Bekenstein,
Phys.\ Rev.\ D {\bf 7} (1973) 2333.
%
\bibitem{Casadio:2021cbv}
R.~Casadio,
Eur. Phys. J. C \textbf{82} (2022) 10
[arXiv:2103.14582 [gr-qc]].
%
\bibitem{Casadio:2023ymt}
R.~Casadio,
Phys. Lett. B \textbf{843} (2023) 138055
[arXiv:2304.06816 [gr-qc]].
%
\bibitem{Misner:1964je}
C.~W.~Misner and D.~H.~Sharp,
Phys. Rev. \textbf{136} (1964), B571.
%
\bibitem{Hernandez:1966zia}
W.~C.~Hernandez and C.~W.~Misner,
Astrophys. J. \textbf{143} (1966) 452.
%
\bibitem{Casadio:2021eio}
R.~Casadio,
Int. J. Mod. Phys. D \textbf{31} (2022) 2250128
[arXiv:2103.00183 [gr-qc]].
%
\bibitem{Casadio:2022ndh}
R.~Casadio, A.~Giusti and J.~Ovalle,
Phys. Rev. D \textbf{105} (2022) 124026
[arXiv:2203.03252 [gr-qc]].
%
\bibitem{Casadio:2023iqt}
R.~Casadio, A.~Giusti and J.~Ovalle,
JHEP \textbf{05} (2023) 118
[arXiv:2303.02713 [gr-qc]].
%
\bibitem{Lukash:2013ts}
V.~N.~Lukash and V.~N.~Strokov,
Int. J. Mod. Phys. A \textbf{28} (2013) 1350007
[arXiv:1301.5544 [gr-qc]].
%
\bibitem{Carballo-Rubio:2023mvr}
R.~Carballo-Rubio, F.~Di Filippo, S.~Liberati and M.~Visser,
``Singularity-free gravitational collapse: From regular black holes to horizonless objects,''
[arXiv:2302.00028 [gr-qc]].
%
\bibitem{Makela:1997rx}
J.~Makela and P.~Repo,
Phys. Rev. D \textbf{57} (1998) 4899
[arXiv:gr-qc/9708029 [gr-qc]].
%
\bibitem{Berezin:1996ju}
V.~A.~Berezin,
Phys. Rev. D \textbf{55} (1997) 2139
[arXiv:gr-qc/9602020 [gr-qc]].
%
\bibitem{Casadio:2013tma}
R.~Casadio,
Eur. Phys. J. Plus \textbf{139} (2024) 770
[arXiv:1305.3195 [gr-qc]].
%
\bibitem{Casadio:2013aua}
R.~Casadio and F.~Scardigli,
Eur. Phys. J. C \textbf{74} (2014)  2685
[arXiv:1306.5298 [gr-qc]].
%
\bibitem{Calmet:2015pea}
X.~Calmet and R.~Casadio,
Eur. Phys. J. C \textbf{75} (2015) 445
[arXiv:1509.02055 [hep-th]].
%
\bibitem{Casadio:2015rwa}
R.~Casadio, O.~Micu and D.~Stojkovic,
JHEP \textbf{05} (2015) 096
[arXiv:1503.01888 [gr-qc]].
%
\bibitem{Feng:2024nvv}
W.~Feng, A.~Giusti and R.~Casadio,
``Horizon quantum mechanics for coherent quantum black holes,''
[arXiv:2408.17091 [gr-qc]].
%
\bibitem{Arrechea:2024nlp}
J.~Arrechea, S.~Liberati and V.~Vellucci,
``Whispers from the quantum core: the ringdown of semiclassical stars,''
[arXiv:2407.08807 [gr-qc]].
%
\end{thebibliography}
\end{document}